# A universal fractal structure of 2d quantum gravity for $c > 1$

*Jan Ambjørn* and *Gudmar Thorleifsson*

The Niels Bohr Institute

Blegdamsvej 17, DK-2100 Copenhagen Ø, Denmark

## Abstract

We investigate the fractal structure of $2d$ quantum gravity coupled to matter by measuring the distributions of so-called baby universes. We demonstrate that the method works well as long as $c \leq 1$. For $c > 1$ it is not clear what distribution to expect. However, we observe strikingly similar distributions for various kinds of matter fields with the same $c$. This indicate that there might be some range of $c > 1$ where the central charge of the matter fields alone determines the fractal structure of gravity coupled to matter. The hypothesis that the string susceptibility $\gamma = 1/3$ is found to be compatible with the data for $1 < c \leq 4$.

# 1 Introduction

The fractal and self-similar structure of $2d$ quantum gravity depends on the central charge of the matter fields coupled to gravity. For $c < 1$ the relation is given by the famous KPZ formula [1]

$$\gamma = \frac{c - 1 - \sqrt{(1-c)(25-c)}}{12}, \qquad (1)$$

where $\gamma$ denotes the entropy (or string susceptibility) exponent of the theory. The detailed relationship of $\gamma$ to the fractal structure of the ensemble of random surfaces dictated by the partition function was analyzed in a recent paper by Jain and Mathur. $\gamma$ determines the distribution of so-called baby universes which can branch off from the "parent universe". In [3] it was shown that the measurement of baby universe distributions is an efficient way to determine $\gamma$ from numerical simulations. In this letter we report the results from extensive simulations using this method to determine the fractal structure of $2d$ gravity coupled to matter fields.

# 2 The method

Let us briefly summarize the method: We discretize the theory of $2d$ gravity coupled to matter by the use of dynamical triangulations [4, 5, 6]. The partition function can be written as

$$Z(\mu, \beta) = \sum_N Z(N, \beta) e^{-\mu N}, \qquad (2)$$

where $N$ denotes the number of triangles, $\mu$ the cosmological constant and $\beta$ some matter field coupling. Critical behaviour for $\mu \to \mu_c$ is obtained if $Z(N, \beta)$ has the following asymptotic behaviour for large $N$:

$$Z(N, \beta) \sim e^{\mu_c(\beta) N} \, N^{\gamma(\beta)-3} \log^{\alpha(\beta)} N. \qquad (3)$$

We have in (3) allowed for a logarithmic correction term of the type that is known to play a role for $c = 1$.

We assume spherical topology and define for a given triangulation a (minimum neck) baby universe as follows: If we cut open the triangulated surface consisting of $N$ triangles along a closed loop made of three different links it will be separated in two disconnected parts, consisting of $B$ and $N - B$ triangles, where $B \leq N/2$. We call the part of triangulation containing the $B$ triangles the baby universe. From (3) it follows that the average number of baby universes consisting of $B$ triangles on a closed triangulated surface of spherical topology made of $N$ triangles is given by

$$\bar{n}_N(B) \sim [(N-B)B]^{\gamma(\beta)-2} \, [\log(N-B) \log B]^{\alpha(\beta)}. \tag{4}$$

It is now clear that a measurement of $\bar{n}_N(B)$ for a fixed large $N$ in principle should be sufficient to determine both $\gamma(\beta)$ and $\alpha(\beta)$.

The matter fields we couple to 2d gravity will be a variety of (multiple) $q$-state Potts models, $q = 2, 3, 4$, all with $0 < c \leq 4$. The matter fields are placed in the center of the triangles and the implementation is standard (for details we refer for instance to [7]). In the case of the $q$-state models we have a coupling constant $\beta$ multiplying the matter part of the action. If we have a multiple $q$-state model we still consider only the case where a single common coupling constant multiplies the different identical copies in the action. We used canonical Monte Carlo simulations, and lattice sizes ranging from 1000 to 2000 triangles. To update the spin configurations a Swendson-Wang cluster algorithm was used, while for the graphs we used a standard link-flip algorithm. Each Monte Carlo sweep consisted of one cluster update and $N$ link flips. Usually $2.5 \times 10^6$ sweeps where performed, but close to $\beta_c$ we used up to $3 \times 10^7$ sweeps.

For $c \leq 1$ we know that the models in the infinite volume limit ($\mu = \mu_c$) undergo a phase transition from a phase with no magnetization to one with constant magnetization for a certain value $\beta_c$ of $\beta$. At this point $\gamma(\beta)$ changes from its value without coupling to gravity to a new value, given by (1). This gives us non-trivial tests of our method. On a finite lattice the jump from the pure gravity cannot take place at a single $\beta$ value, and we see a peak around the critical value. However, as will be clear from the numerical results, the peak is very sharp.

For $c > 1$ we have no analytical results. Numerical studies suggest that we will still observe a transition between a phase with zero total magnetization and a magnetized phase. However, as reported in [7] the picture seems somewhat more complicated if $c$ is large since the back-reaction of the matter fields on gravity was largest *before* the transition, i.e. for $\beta < \beta_c$. We will return to a discussion of these results in sec. 4.

## 3   The measurements

We have measured $\bar{n}_N(B)$ with high statistics and for a wide range of couplings in the following models: single $q = 2, 3, 4$-state Potts model ($c = 1/2$, $4/5$ and $1$), a double $q = 3$-state Potts model ($c = 8/5$) and a quadruple $q = 3$-state Potts model ($c = 16/5$). We also looked at double $q = 4$-state and quadruple $q = 2$-state models ($c = 2$). and a quadruple $q = 4$-state model ($c = 4$), for couplings around $\beta_c$. The measured distributions $\bar{n}_N(B)$ are fitted to

$$\log \bar{n}_N(B) = A + (\gamma - 2) \log \left[ B(1 - \frac{B}{N}) \right] + \alpha \log \left[ \log B \log(N - B) \right] + \frac{C}{B}. \quad (5)$$

The last term is a "phenomenological" finite size correction term. The asymptotic form (3) of $Z(N, \beta)$ is only assumed valid for $B \gg 1$. A simple first correction could be of the form

$$B^{\gamma-2} \to B^{\gamma-2} \left( 1 + \frac{\tilde{C}}{B} + \cdot \right). \quad (6)$$

It leads to the last term in (5), and has been shown to improve the results for small $B$ considerable [3].

In fig. 1 we show the results of the fits for $\gamma$, with and without the additional parameter $\alpha$ from the logarithmic corrections, for values of $\beta$ near the critical value $\beta_c$ for the various models. The inclusion of $\alpha$ in the fit increases the errorbars on $\gamma$ for the obvious reason that it can be hard to distinguish the two contributions over a limited interval of $B$.

The general picture is as expected: we see a peak in $\gamma(\beta)$ near $\beta_c$, while $\gamma(\beta)$ has essentially reached the $c = 0$ value $\gamma(c = 0) = -1/2$ as soon as $|\beta - \beta_c| > 0.1$. This is a clear indication that $\gamma$ changes at the critical point. Naively one would expect the peak to get more narrow if we increase the number of triangles $N$ used in the simulation, but we have only seen a weak dependence on $N$. If we compare the functions $\gamma(\beta)$ for $c < 1$ with the corresponding function for $c = 1$ we see a marked difference in the shape. For $c < 1$ there is a very narrow and pronounced top in the peak as already mentioned, while the peak is rather broad for $c = 1$. Further the peak value of $\gamma(\beta)$ gives the correct result for $c < 1$ when we do not include logarithmic corrections. If we include logarithmic corrections the value will be somewhat to high. Since we know from analytical results that there are no logarithmic corrections for $c < 1$ we conclude that it is difficult to include $\alpha$ as a free parameter in the fit. For $c = 1$ the situation is different. The fit without logarithmic correction gives too small a peak value for $\gamma(\beta)$. If fact the value is *smaller* than the corresponding values for $c = 1/2$ and $c = 4/5$. The very sharp peaks of the $c < 1$ curves are simply missing. Including the logarithmic correction on the other hand gives the expected value of $\gamma$ at the peak ($\gamma = 0.05 \pm 0.08$ for $N = 1000$). The logarithmic exponent was harder to determine but we got $\alpha = -1.6 \pm 0.3$ (compared to the expected value of $\alpha(\beta_c) = -2$).

Our conclusion is that the baby universe technique for extracting $\gamma$ works very well for $c \leq 1$, and that triangulations of the size 1000-2000 triangles and baby universe distributions of a size from 10-500 are sufficient to reflect in a precise way the fractal structure of $2d$ quantum gravity coupled to matter with $c \leq 1$,. However, this conclusion is based on the knowledge that there are no logarithmic correction for $c < 1$.

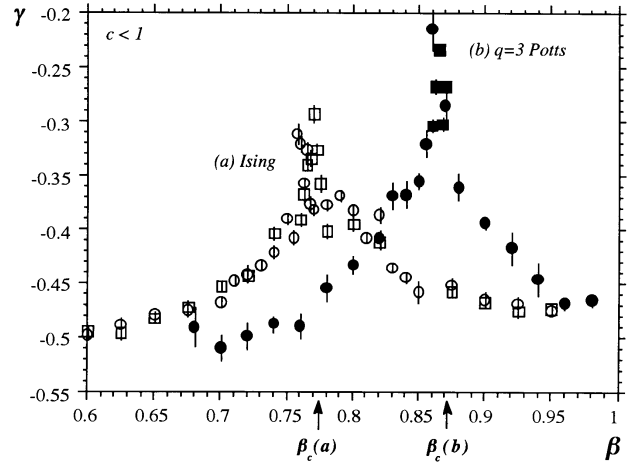

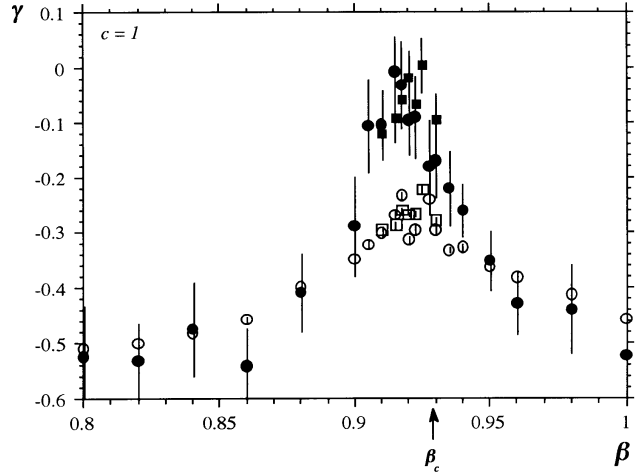

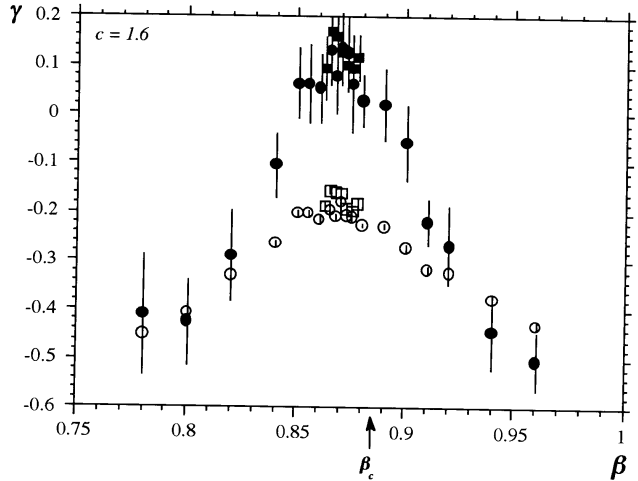

Figure 1: Measured value of $\gamma$. In all figures circles denote lattice size $N = 1000$ and squares $N = 2000$. In fig. 1a we shown 1 Ising model (open symbols) and 1 $q = 3$-state Potts model (filled symbols). The fits are without logarithmic corrections. In fig. 1b we show 1 $q = 4$-state Potts model and in fig. 1c 2 $q = 3$-state Potts models. Open symbols show fits without logarithmic correction and filled symbols fits with logarithmic corrections.

|   |          | Fit $a$       |          | Fit $b$      |            |          | Fit $c$    |          |
|---|----------|---------------|----------|--------------|------------|----------|------------|----------|
| c | Model    | $\gamma$      | $\chi^2$ | $\gamma$     | $\alpha$   | $\chi^2$ | $\alpha$   | $\chi^2$ |
| 1 | 1 Gaussian | -0.334(12)  | 172      | -0.05(8)     | -1.60(30)  | 213      | -3.55(13)  | 339      |
|   | 1 $q=4$ P. | -0.313(15)  | 206      | -0.020(50)   | -1.56(13)  | 197      | -3.45(10)  | 325      |
| 1.6 | 2 $q=3$ P. | -0.212(15) | 287      | 0.111(70)    | -1.78(21)  | 220      | -2.97(10)  | 264      |
| 2 | 2 Gaussian | -0.152(12)  | 300      | 0.161(64)    | -1.77(28)  | 236      | -2.67(13)  | 263      |
|   | 2 $q=4$ P. | -0.149(8)   | 383      | 0.202(34)    | -1.94(17)  | 264      | -2.76(10)  | 309      |
|   | 4 $q=2$ P. | -0.167(8)   | 283      | 0.157(50)    | -1.75(30)  | 214      | -2.79(13)  | 209      |
| 3 | 3 Gaussian | -0.010(10)  | 526      | 0.53(11)     | -3.01(47)  | 227      | -1.95(13)  | 246      |
| 3.2 | 4 $q=3$ P. | 0.024(14)  | 474      | 0.56(15)     | -3.09(27)  | 205      | -1.87(20)  | 220      |
| 4 | 4 Gaussian | 0.079(12)   | 584      | 0.60(8)      | -3.05(37)  | 223      | -1.47(10)  | 241      |
|   | 4 $q=4$ P. | 0.084(10)   | 956      | 0.64(9)      | -3.16(92)  | 180      | -1.51(10)  | 256      |
| 5 | 5 Gaussian | 0.217(10)   | 1374     | 0.97(17)     | -4.34(75)  | 155      | -0.93(10)  | 358      |

Table 1: Exponents and $\chi^2$ values for fits to $\log \bar{n}_N(B)$ for $c \geq 1e$ with $B \geq 11$ and 245 data points. The following functional forms are used:
(a) $A + (\gamma - 2)\log[B(1 - B/N)] + C/B$
(b) $A + (\gamma - 2)\log[B(1 - B/N)] + \alpha \log[\log B \log(N - B)] + C/B$
(c) $A + (1/3 - 2)\log[B(1 - B/N)] + \alpha \log[\log B \log(N - B)] + C/B$

Let us now turn to a discussion of the results for $c > 1$. It has often been argued that $c > 1$ has a lot in common with gravity in higher dimensions. The program of dynamical triangulation has been formulated and implemented in higher dimensions [9], and one of the lessons is that we cannot take the functional form (3) for granted. It seems that the generic form is rather like [10]

$$Z(N) \sim e^{\mu_c N - \lambda N^\alpha}, \qquad \alpha < 1. \tag{7}$$

We have tested whether $Z(N, \beta)$ for $c > 1$ should fit better to such a functional form than to (3), but found that it is not the case. Even though in some cases the $\chi^2$ of the fits where not bad, the fitted parameters had absurd values. As a consequence we will assume that (3) is a reasonable functional form. However, for $c > 1$ we have no a priori knowledge of the logarithmic corrections. It is notoriously difficult to distinguish between logarithmic and small power law corrections, but the good results for the $c = 1$ case might give some hope that one can extract a reliable the value for $\gamma$. As will be obvious from table 1 and sec. 4 this will not be the case[1]. The trend of broader peaks and larger values of $\alpha$ if extracted from (3) continues with increasing $c$. Again one could be worried that we use too small triangulations. To test this we have performed a simulation with two $q = 3$ models coupled to gravity and $N = 100.000$. This allowed us to measure baby universes with reasonable statistics up to a size of 5000, i.e. a factor 10 more for the smaller surfaces (N= 1000 and N= 2000) used above. We saw however no drastic change in the extracted value of $\gamma$. If extracted with $\alpha$ set to zero, $\gamma$ changed from $-0.212(15)$ to $-0.175(24)$. If extracted with logarithmic correction the numbers were $0.11(7)$ and $0.19(9)$,respectively. In both cases the $\chi^2$ of the fits were acceptable and we have no numerical reason for preferring one kind of ansatz from the other.

The results obtained for $c \geq 1$ are shown in table 1 together with results obtained for Gaussian models [3]. For the same $c$ we get the same $\gamma$'s and, within the precision available, the same $\alpha$'s[2]. One is lead to the hypothesis that there might be universality even for $c > 1$, at least in some region $1 < c < c_0$ (the largest value of $c$ where we have compared distributions in detail is $c = 4$).

This becomes even more striking when one makes a direct comparison between the distributions $\bar{n}_N(B)$ of baby universes for $c = 1, 2$ and 4. In fig. 2 we show the measured distributions (normalized by the the distribution for pure gravity) for the following models: For $c = 1$ 1 Gaussian and 1 $q = 4$-state Potts model, for $c = 2$ 2 Gaussian, 2 $q = 4$-state Potts and 4 Ising models, and for $c = 4$ 4 Gaussian and 4 $q = 4$-state Potts models. The advantage of this representation is that it is independent of any fits and assumptions about the functional form of the distributions. *We see a pronounced universality in the curves for fixed central charge.* It seems as if the back-reaction of matter on gravity only depends on the central charge even if $c > 1$.

In order to make the claim more substantial we have statistically tested the hypothesis that the distributions are identical, module a possible non-universal normalization and phenomenological correction terms. So we fitted the difference between two distributions $\bar{n}_{1_N}$ and $\bar{n}_{2_N}$, with the same $c$, to the form

$$\log \bar{n}_{1_N}(B) - \log \bar{n}_{2_N}(B) = a_1 + \frac{a_2}{B}, \qquad (8)$$

and tested the goodness of the fits. The fits where made with $B \geq 11$ and 245 data points. In table 2 we show the resulting $\chi^2$ for the fits and the corresponding significance levels $P(\chi^2_{243} \geq \chi^2)$. As can be seen the significance levels are in all cases larger than 50%, so we conclude that the data is consistent with the hypotheses.

---

[1] A similar conclusion as been reached in two recent papers using different observables [11, 12].
[2] By $\gamma$ and $\alpha$ for the spin models we mean the values measured at the (finite size) critical $\beta$ as identified by the peak value of $\gamma(\beta)$. The Gaussian models are alway critical (before coupling to $2d$ gravity) and any coupling constant can be absorbed in a redefinition of the cosmological constant.

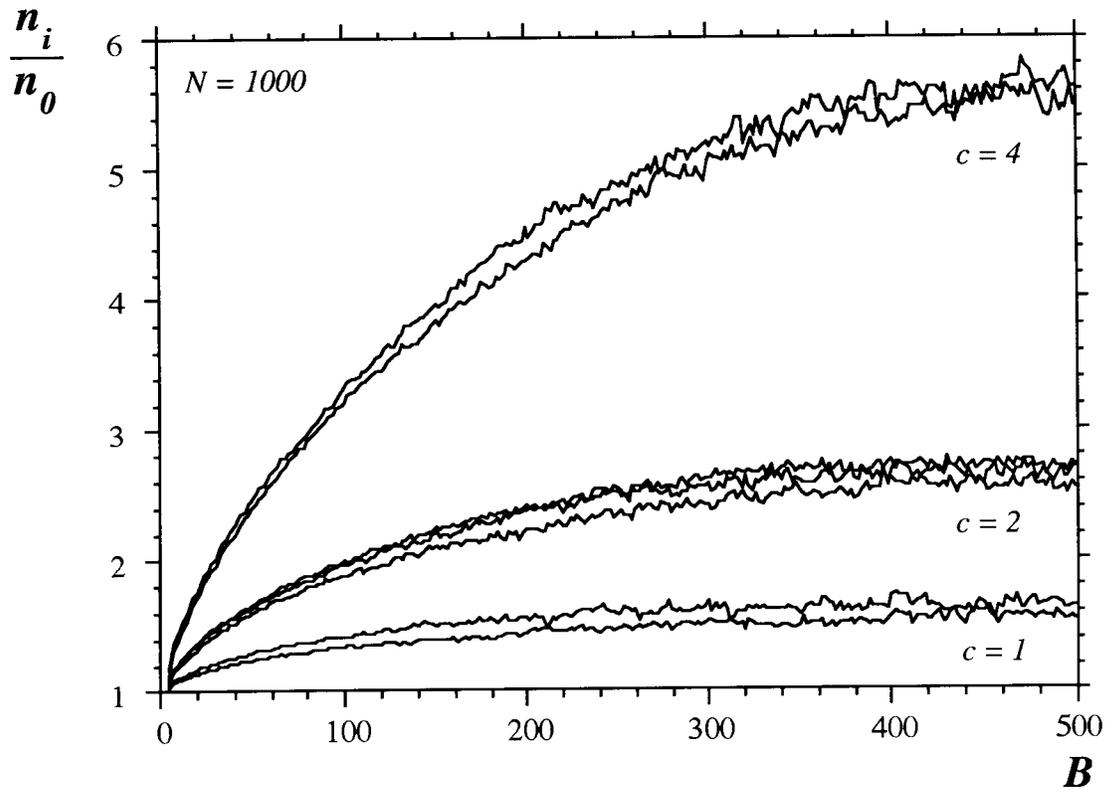

Figure 2: Normalized distributions of baby universes for various matter fields coupled to $2d$ gravity.

| $c$ | $\bar{n}_1$ | $\bar{n}_2$ | $\chi^2$ | $P(\chi^2_{243} \geq \chi^2)$ |
|---|---|---|---|---|
| 1 | 1 Gaussian | 1 $q=4$ Potts | 252 | $>50\%$ |
| 2 | 2 Gaussian | 2 $q=4$ Potts | 201 | $>95\%$ |
| 2 | 2 Gaussian | 4 $q=2$ Potts | 228 | $>75\%$ |
| 4 | 4 Gaussian | 4 $q=4$ Potts | 248 | $>50\%$ |

Table 2: Quality of the fits to (8) for comparing distributions of baby universes with same central charge

Contrary to the observables used in [7, 8], where Gaussian models and multiple Potts models coupled to $2d$ gravity were also compared, and which showed a surprising universality between various Potts models with the same central charge, but some deviation between the Potts models and Gaussian models, the distributions of baby universes should (for large $B$) be characterized only by their critical exponents. They are consequently much better observables when it comes to a comparison between different models.

## 4 The quest for a $c > 1$ hypothesis

The results reported above show that distributions of baby universes provide us with a convenient tool for analyzing the interaction between matter and $2d$ gravity. They also suggest that the fractal structure of the random surfaces depends only on the central charge of matter, at least in the range $1 < c < 4$ we have tested by numerical simulations. Unfortunately the method is unable, with the present quality of the numerical results, to nail down the correct values of $\gamma$ and $\alpha$ in (3) as they correspond to terms with too similar functional behavior and a change in one can be absorbed in the other without seriously affecting the quality of the fit. From this point of view it would be very interesting to have an educated guess of a more restricted functional form than the one given by (3). Let us end by a brief discussion of some recent speculations in that direction [13, 14, 15].

It has been argued that $\gamma > 0$ implies $\gamma = 1/(n+1)$, $n > 1$, [13, 14]. Specific matrix models which realize $\gamma = 1/(n+1)$ are known [17]. For spin systems it is natural to imagine that $\gamma = 1/3$ is realized, at least for multiple spin systems with $c \gg 1$. The argument goes as follows: The value $\gamma = 1/(n+1)$ reflects a fractal structure of the surfaces where the surface is glued together of "bubbles" of $\bar{\gamma} = -1/n$ matter. The results in [7] strongly suggest that the individual bubbles are already magnetized in some range of $\beta < \beta_c$ and that the phase transition to a magnetized phase is just an alignment of the spins in the individual bubbles. This would imply $\bar{\gamma} = -1/2$ (pure gravity) and therefore

$$\gamma = \frac{1}{3}. \qquad (9)$$

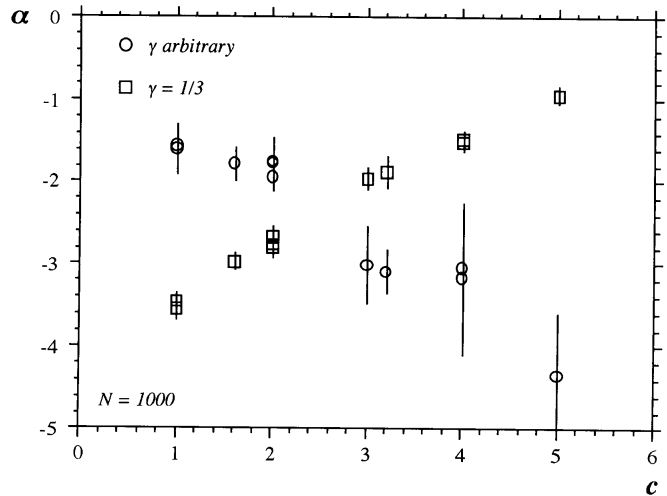

Figure 3: The logarithmic correction exponent $\alpha$ for the fits (5) and (10).

An identical scenario was obtained independently by Wexler using mean field theory [15]. While this result might well be correct for multiple Potts models with $c \gg 1$ it is not clear that it is valid for Gaussian models for $c \gg 1$, where other mean field results suggest that $\gamma = 1/2$ or $\gamma = -c/2$, depending on the weight used for the triangulations [16]. However, it is tempting to conjecture that $\gamma = 1/3$ should be valid all the way down to $c = 1$ for multiple Potts models, and since we have seen that the baby universe distributions (within the numerical accuracy) for small $c > 1$ are the same for Gaussian and Potts models, there should be a range $1 < c < c_0$ where the Gaussian models also have $\gamma = 1/3$. If we combine this conjecture with the numerical observation that the baby universe distributions change continuously in some range $c \geq 1$, we are lead to the suggestion

$$Z_c(N) = e^{\mu_c N} N^{1/3-3} \log^{\alpha(c)} N, \qquad c > 1, \qquad \alpha(c) \to -\infty \text{ for } c \to 1. \qquad (10)$$

We have tested whether this conjecture is in accordance with our baby universe measurements for $c > 1$. The result is shown in table 1, where we show extracted values of $\alpha$ and the quality of the fits. We see that $\chi^2$ of fits made with the assumption $\gamma = 1/3$ and with $\gamma$ as a free parameter is comparable. This illustrates further that additional information is needed if we want to have some hope of extracting reliable values for $\gamma$. It also shows that the conjecture (10) cannot be ruled out. It is interesting to note that we get a very different functional dependence $\alpha(c)$ for the logarithmic exponent $\alpha$ when we compare the results where $\gamma$ is fixed to 1/3 and results where $\gamma$ is arbitrary. This is shown in fig. 3, and we observe that the functional form $\alpha(c)$ for $\gamma = 1/3$ indeed is compatible with (10), except that we cannot expect $\alpha(c) \to -\infty$ for finite size lattices.